# DEVELOPMENT AND EVALUATION OF RECURRENT NEURAL NETWORK BASED MODELS FOR HOURLY TRAFFIC VOLUME AND AADT PREDICTION

**Zadid Khan, Corresponding Author**
Ph.D. Student
Clemson University
Glenn Department of Civil Engineering
351 Fluor Daniel Engineering Innovation Building, Clemson, SC 29634
Tel: (864) 359-7276, Fax: (864) 656-2670
Email: mdzadik@clemson.edu

**Sakib Mahmud Khan**
Ph.D. Candidate
Glenn Department of Civil Engineering, Clemson University
351 Fluor Daniel Engineering Innovation Building, Clemson, SC 29634
Tel: (864) 569-1082, Fax: (864) 656-2670
Email: sakibk@clemson.edu

**Kakan Dey, Ph.D.**
Assistant Professor
Department of Civil and Environmental Engineering
West Virginia University
ESB 647, 401 Evansdale Drive, PO Box 6070, Morgantown, West Virginia
Tel: (304) 293-9952   Fax: (864) 656-2670
Email: kakan.dey@mail.wvu.edu

**Mashrur Chowdhury, Ph.D., P.E., F.ASCE**
Eugene Douglas Mays Endowed Professor of Transportation and
Professor of Automotive Engineering
Clemson University
Glenn Department of Civil Engineering
216 Lowry Hall, Clemson, South Carolina 29634
Tel: (864) 656-3313   Fax: (864) 656-2670
Email: mac@clemson.edu

Word count: Abstract = 244 words, Text 7025 + 5 Tables 1250 + 17 Figures = 8275 words, Reference = 712 words

**ABSTRACT**
The prediction of high-resolution hourly traffic volumes of a given roadway is essential for transportation planning. Traditionally, Automatic Traffic Recorders (ATR) are used to collect this hourly volume data. These large datasets are time series data characterized by long-term temporal dependencies and missing values. Regarding the temporal dependencies, all roadways are characterized by seasonal variations that can be weekly, monthly or yearly, depending on the cause of the variation. Regarding the missing data in a time-series sequence, traditional time series forecasting models perform poorly under the influence of seasonal variations. To address this limitation, robust, Recurrent Neural Network (RNN) based, multi-step ahead forecasting models are developed for time-series in this study. The simple RNN, the Gated Recurrent Unit (GRU) and the Long Short-Term Memory (LSTM) units are used to develop the model and evaluate its performance. Two approaches are used to address the missing value issue: masking and imputation, in conjunction with the RNN models. Six different imputation algorithms are then used to identify the best model. The analysis indicates that the LSTM model performs better than simple RNN and GRU models, and imputation performs better than masking to predict future traffic volume. Based on analysis using 92 ATRs, the LSTM-Median model is deemed the best model in all scenarios for hourly traffic volume and AADT prediction, with an average RMSE of 274 and MAPE of 18.91% for hourly traffic volume prediction and average RMSE of 824 and MAPE of 2.10% for AADT prediction.

## INTRODUCTION

The hourly traffic volume roadway data is an important high-resolution dataset used to describe the operational characteristics of a transportation system. Accurate hourly traffic volumes can be utilized in calculating the Average Annual Daily Traffic (AADT). AADT is an essential parameter in many transportation models and decisions. Moreover, the prediction of future hourly traffic volumes of a given roadway is even more important than current data because it can be used to estimate future growth. Here, the volume growth factor of a roadway can be combined with other external data to predict the overall growth pattern of an area. Moreover, the high-resolution data provides insight into the factors contributing to growth, as it may be a gradual growth pattern or a sudden peak. The roadway volume can increase at a very specific time next year due to some special events, and a predictive model can predict this change. This means that the special event is a phenomenon that has occurred before. However, if a predictive model is unable to capture this event, then it is a new phenomenon that has not been previously observed. Therefore, the detection of special events or anomalies is also an application of high-resolution hourly volumes.

Transportation planning is characterized by many projects that are related to future infrastructure investments (i.e., designing new roadways and bridges, urbanization/land development, the addition of new lanes in a roadway, new medians, and new traffic signals). As such, the expensive and time-consuming nature of these projects means that the project team must ensure the importance of a project before an investment is forthcoming. A predictive model that accurately forecasts hourly volumes and AADT can help in increasing the confidence in the investment decisions. Moreover, many transportation software and models use AADT as input data, which, if the future AADT value is reliable, can be used to inform a predictive analysis. For example, AADT serves as input into the Safety Analyst Software through which city planners can analyze and improve traffic safety (*1*).

There are several ways to collect continuous hourly traffic volumes for a roadway, most notably using Automatic Traffic Recorders (ATRs), which are permanent traffic count stations strategically located at selected locations. These ATRs collect hourly volumes and can collect real-time average speed data. The technology used in the ATR can vary according to data collection requirements. The most popular technologies used in ATR are inductive loops, magnetic counters, radar sensors and surveillance camera. The ATRs usually send the real-time data to a Department of Transportation (DOT) traffic data center, where it is archived and made available for internal and external usage through an open portal. State DOTs also collect short-term counts at hundreds of locations to supplement the permanent count station data.

The hourly traffic volume data from an ATR is a classic example of a time series with multiple sub-patterns. The hourly volume counts for a day exhibits hourly count variation by the time of the day, as the volumes at off-peak hours (e.g., nighttime) are always less than during the day. Weekday and weekend variations also characterize such systems, with weekdays subject to a higher volume than weekends, or vice versa, depending upon the location. The volumes also vary from month to month depending on weather, seasonal events and other external factors. Different changes to the roadway such as a temporary shutdown of lanes and special events can create random variation in the time series, which may or may not be repeated every year. Finally, the ATR stations can be temporarily unavailable due to faulty equipment, and disruption in communication with the traffic data center.

The objective of this study is prediction of hourly volume for 365 days of the following year so that an accurate estimation of the AADT of the next year can be derived from the data of the previous years. Moreover, given that the high-resolution hourly volume prediction should capture all the hourly, weekly and monthly variations that are present in the dataset from previous years, a predictive model is needed for multi-step-ahead forecasting.

In this study, we detail the development of such a model, the Recurrent Neural Network (RNN) based time-series forecasting model with missing data treatment for predicting hourly traffic volume and AADT. RNN models are specialized deep learning models for sequence prediction that capture the different variation patterns in previous time steps and make a reliable estimate of future time steps (*2*). Unfortunately, the missing data is a major problem in time series because they are part of the sequence and cannot be discarded from the dataset. Moreover, hourly volumes for every hour in the input dataset are required to calculate an accurate estimate of AADT from the ATR data. Given that the removal of missing data will greatly reduce the prediction accuracy, a solution for this missing data input is required. In this paper, we describe the two approaches used, combined with the RNN, to address this issue. The architecture of the general predictive model is presented, including the missing data treatment layer and the RNN layer. We evaluate different variations of the proposed model to identify the best model accordingly.

## LITERATURE REVIEW
Based on the context of this research, the literature review part has been divided into three segments;
- Missing data in prediction;
- Time series prediction with RNN;
- Traffic volume and AADT prediction.

### Missing data in prediction
An incomplete dataset is always a challenge for time-series forecasting. However, imputing missing values in the historical data can provide a more robust prediction model. Comparative studies have been performed on various imputation methods for traffic data (*3*) and health survey data (*4*). The authors in (*5*) develop GRU-D from the basic GRU concept to mask the missing data in the historical database. They use the prediction model on synthetic and real-world clinical datasets. The GRU-D model exhibited a similar complexity and computation time compared to original RNN models. The authors also compared with basic RNN models coupled with other baseline imputation methods such as Expectation Maximization, Principal Component Analysis, K-nearest neighbor, and Softimpute. The authors found that GRU-D model performs better than other imputation methods. Similarly, in their use of pattern mixture kernel submodels (PMKS), which includes submodels for each missing data pattern, the authors in (*6*) found that PMKS outperforms different imputation models and complete-case submodels.

### Time Series Prediction with RNN
Time series analysis is a major field of research. The state of the art approaches to time series prediction has been discussed in (*7*). Deep learning models have been applied to traffic data prediction and it has revealed their potential to make high accuracy prediction. The authors in (*8*) conduct a comparative study about unsupervised learning and deep learning models for time series prediction. RNN model is one of the popular methods for sequence prediction. Internal feedback connections with the hidden RNN layers have been used to model the temporal relationship within the time-series data explicitly. Based on the actual output and the predicted output, the error is calculated, and the weight of the networks are revised until the model converges. RNN, which is also notable for a high prediction accuracy for noisy datasets, has been successfully used in financial forecasting (*9*), health care (*5*), transportation, chaotic time series prediction (*10*), and acoustic modeling (*11*). In that regard, the authors in (*9*) applied RNN in financial forecasting using financial datasets that are small, noisy, and non-linear. Using RNN, they combined symbolic methods with RNN to overcome the issues related to the unequal a-priori class probabilities and overfitting. Also, by rejecting data having low confidence measure, the prediction model achieved 40% less error. In their use of Long Short-Term Memory (LSTM) to predict a large scale acoustic model, the authors in (*11*) noted that the five-layered LSTM RNN outperformed one, two, three and seven-layered LSTM RNN networks. These networks were trained with a hand-transcribed and anonymized dataset with three million distinct pieces of data.

For multi-step prediction (i.e., predicting for a time series sequence), the authors in (*10*) used RNN where nodes are operated non-linearly, the output of which was linked with the input of that node and the following node. A self-adaptive and extended Kalman filter based back propagation method was used to achieve a superior level of performance of the extended method (i.e., Root Mean Squared Error (RMSE) of 3.9 for 6 min lag time) compared to the standard RNN model (i.e., RMSE 4.4 for a 6 min lag time). However, the increase of time horizon for the multistep prediction caused a decrease in the prediction accuracy of all models. The authors in (*12*) reported similar findings regarding the multi-stage predicted value of the current time step serving as the input for the next time step. A single-step RNN model outperformed the multi-stage RNN model, the linear regression model, and the hidden Markov model.

**Traffic volume and AADT prediction**

Previous research work has been successful in applying deep learning to traffic flow prediction problem. The authors in (*13*) have developed a stacked autoencoder based model, while the authors in (*14*) have developed deep belief networks with multitask learning. For predicting short-term traffic flow, the authors in (*15*) used an LSTM model with dynamically optimal time lag to predict short term traffic. To determine the optimal time lag in real-time, the LSTM RNN was used with memory blocks of three multiplicative units. Thirty loop detection datasets from six California freeways were used (from California PeMS Database). The data was aggregated at 15, 30, 45 and 60-minute intervals. The dynamic time lag was then used to derive the lowest mean absolute percentage error (MAPE) of 6.5% compared to the support vector machine, the random walk, the single layer feed forward neural network and the stacked autoencoder models. Similarly, in (*16*), LSTM (Mean Absolute Error or MAE 18%) and the Gated Recurrent Units (GRU) neural network (MAE 17%) are applied to California PeMS data to predict an improved short-term traffic flow over the autoregressive integrated moving average (ARIMA) model (MAE 19%). To predict AADT, the authors in (*17*) used support vector regression (SVR), Holt's exponential smoothing (HES) and ordinary least-square linear regression (OLS-regression). Specifically, in their use of a 20-year AADT dataset from Tennessee, they noted that SVR (MAPE 2.3%) outperformed both the Holt-ES (MAPE 2.7%) and OLS-regression (MAPE 3.9%). Another study conducted by the Idaho Department of Transportation noted that the Classification and Regression Trees (CART) method was more accurate than growth factor, linear regression and multiple regression methods (*18*).

**METHOD**

In this section, we will outline the three-step research method as shown in Figure 1, each of which is described in a separate sub-section.

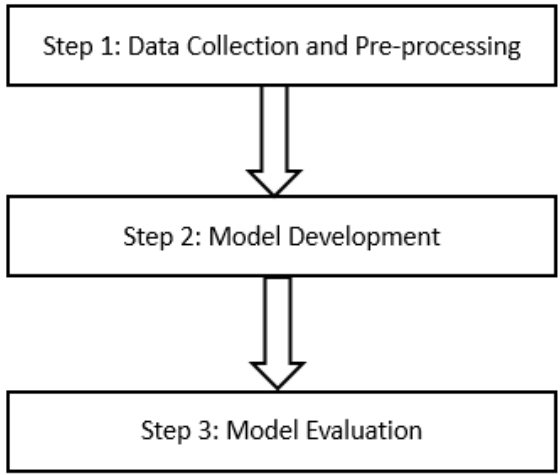

**FIGURE 1 Steps of the research method**

**Step 1: Data Collection and Pre-processing**
Before developing the model, ATR data from 166 ATRs at different locations across the state of South Carolina is collected for past 10 years (2008-2017). We then classify this data into seven functional ATR highway groups, the names and the numbers of which are provided below.
   a.  Rural Interstate (38 ATRs)
   b.  Urban Interstate (45 ATRs)
   c.  Rural Arterial (37 ATRs)
   d.  Urban Arterial (22 ATRs)
   e.  Rural Collector (9 ATRs)
   f.  Urban Collector (6 ATRs)
   g.  Local Roads (6 ATRs)

Of these functional groups, the last group (Local Roads) only have six ATRs, and all six ATRs were established in 2013, so they only have a partial dataset. Therefore, in this study, we use the first six ATR categories from the first six functional groups. Each ATR has a specific identification number, and we collect data from one ATR of each functional group. Although the data is openly available on the South Carolina DOT website, it was not in a structured data file but rather stored in an interactive web interface for the public. We used Python to create a script to collect and store this web interface data into a text file. Data from all ATRs for ten years are collected, and for each ATR, the hourly volume time series contains 87672 hours of data. However, some ATRs have a percentage of missing value greater than 20% over the 10 year period of 2008-2017. Only 92 ATRs out of the 166 ATRs have missing value less than 20%. Only these ATRs are considered throughout the analysis.

After data collection, it is prepared for model input to meet the goal of predicting the volume time series 365 days (a full year) in advance. Therefore, we perform a forward time shift of the data by 365*24 = 8760 hours and create a second time series. The original time series serves as the input and the shifted time series serves as the output. We augment both data series to form the prediction dataset. Due to the forward shift operation, the number of data points in the prediction dataset is 78912. Both the input and the output time series may contain missing data in the form of 'NaN' (Not a Number) values, which are left unchanged in the dataset. 'NaN' is a numeric data type used to fill empty spaces in a numeric vector.

## Step 2: Model Development

The block diagram of the 365-day ahead hourly volume prediction model is detailed in Figure 2. Each block is described below in detail.

### Data Processing
The two time series are first combined and then separated into the training and testing data sets with an input time series of 78912 hours of data. To predict the AADT for the last two years (2016 and 2017) of the dataset, the total number of days in 2016 and 2017 are 731. Therefore, the total number of hours in the testing set is 731*24, or 17544, thus leaving the training set with 61368 hours of data.

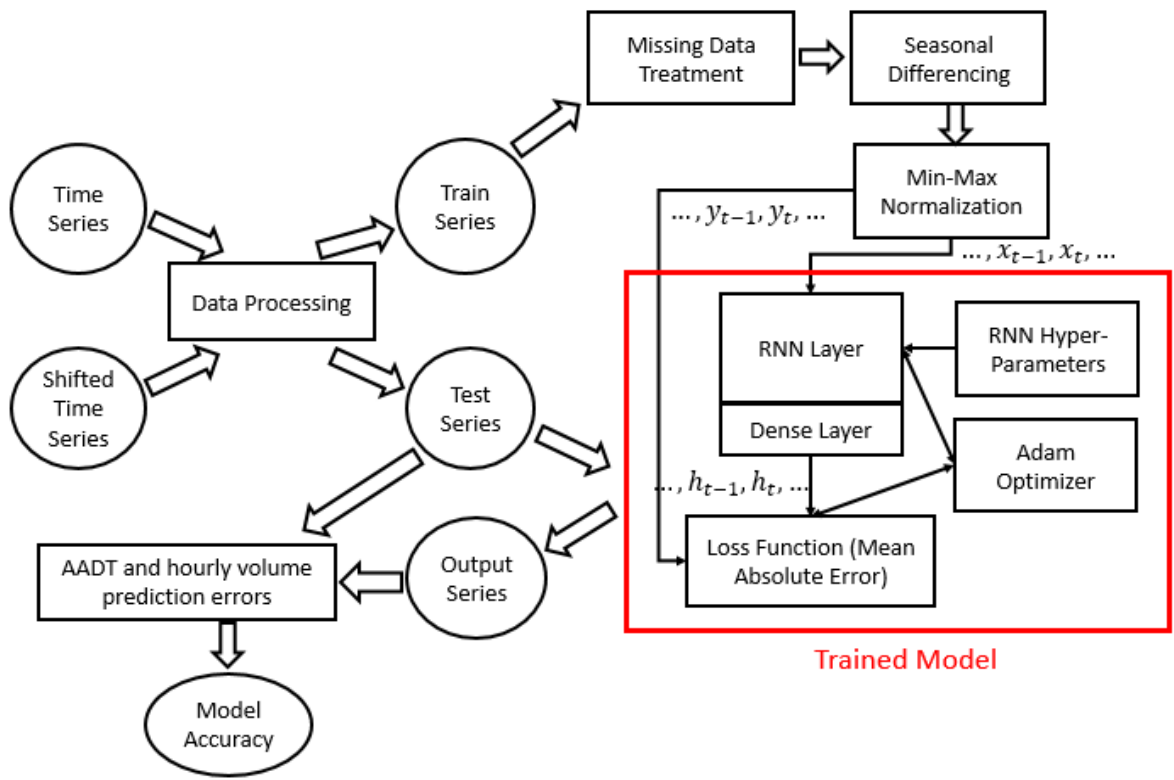

**FIGURE 2 Block Diagram of hourly volume (and AADT) prediction model**

### Missing Data Treatment
To overcome the missing data problem, we use the two approaches of masking and imputation, both of which are described below.

**Masking:** The missing data is masked via a masking layer, the function of which is replacing all missing values with a specific value in the dataset. In the training dataset, the missing values are represented as 'NaN'. The 'NaN' values are then replaced by a masking value, which we set at "-1" here. The model learns that these "-1"s are missing values, so whenever it finds a "-1" in the testing set, it predicts accordingly. The reason for masking with "-1" is to differentiate between actual data and missing data. "-1" will never appear naturally in the hourly volume dataset since this is not a valid hourly volume value. Hence, "-1" is a good masking value.

**Imputation:** The second approach we adopt is data imputation. The function of the imputation layer is estimating the missing values for a variable based on the existing data available for other time-steps and other variables. Although several popular imputation methods are used with time series data, we selected the five below, all of which were implemented in Python.

*Mean/Median Imputation:* This is the simplest form of imputation method, where we separate the dataset from the missing values, take the mean or the median of the separated dataset and replace the missing values with this value in the original dataset. Regarding computation complexity, this is a simple and computationally inexpensive method.

*Expectation Maximization (EM):* This method is used for many different applications, particularly data imputation. The core idea of this algorithm is using the available data to predict the missing data in a dataset (*19*). Let us assume that the set of missing values is $Z$ and the rest of the values form the set $X$. Using the value $X$, we can then develop a model with the parameter set of $\theta$, and then apply that model to estimate the values in $Z$. However, given that both $\theta$ and $Z$ are initially unknown quantities, the iterative process will converge on a solution.

The EM algorithm has two steps, expectation step and maximization step. In the expectation ($E$) step, the algorithm assumes that $\theta$ is known and calculates the expected value of the log-likelihood (LogL) function (Equation 1). The likelihood function calculates a probability value for each point being in a certain subgroup of the overall solution, defined by a subset of the parameters $\theta$. In the maximization ($M$) step, the algorithm finds the optimum parameter that maximizes the log-likelihood function (Equation 2). Maximizing this function ensures that $\theta$ is converging towards the correct values.

$$Q(\theta|\theta^t) = E_{Z|X,\theta^t}[logL(\theta; X, Z)] \tag{1}$$

$$\theta^{t+1} = argmax_\theta \ Q(\theta|\theta^t) \tag{2}$$

Here, $Q$ is the expectation value, $\theta^t$ refers to value of the parameter set $\theta$ at time step $t$.

*Multivariate Imputation by Chained Equations (MICE):* It is a popular method for data imputation. It is used when the dataset has multiple variables. In the first step, all missing values are replaced with a placeholder value (in this study, the mean is used). Afterward, the algorithm assumes one variable as the dependent variable and all other variables as independent variables. The algorithm performs regression (in this study, we have used linear regression) to calculate the missing values for the dependent variables. The same process is performed for all other variables through several cycles until satisfactory results are achieved (*20*).

*K-Nearest Neighbour (KNN):* This is a non-parametric algorithm to impute the missing values in the dataset. The K-nearest neighbor can function for any dataset as long a relationship between adjacent indices of the dataset exists, which is true for the hourly volume time series data. The algorithm calculates the Euclidian distances from the k-nearest neighbor of the missing value and then estimates the missing value as the average of the k-nearest values (*21*). Since its application here is to a time series and the short-term dependency is significant, we choose a small k value of 5 in this study.

*Random Forest (RF):* This algorithm is an extension of the decision tree algorithm. At first, it separates the missing data from the actual dataset. It then randomly samples a subset of data points from the clean dataset during multiple iterations to create numerous decision trees using each subset. Finally, the missing value from all decision trees is calculated to derive the average of the output (*22*).

*Seasonal Differencing and Min-Max Normalization*
After solving the missing data problem, we apply seasonal differencing on the entire training dataset and then apply the Min-Max normalizer. The hourly volume time series for ATRs is a non-stationary time series, which has different trends and seasonal components. A stationary time series is a series whose mean, variance and autocorrelation structures do not change over time. In other words, the time series will not have any time-varying trend or seasonal components. The traffic volumes usually increase over time, and the growth may contain linear or exponential trend or seasonal components. The non-stationary time series as input will result in inaccurate predictions from the RNN model. To avoid this scenario, we have used seasonal differencing on the data so that the time series becomes stationary. The statistical properties remain unchanged for the whole duration. Seasonal differencing can be defined as the process that creates a time series of changes from season to season, usually a year. For seasonal differencing, a lag version of the time series is subtracted from the original time series. We need to identify the interval of time lag for differencing. Since the time series is based on hourly data, we have identified that the optimum interval that works well for all ATRs is 8736. Afterward, we normalize the data between 0 and 1. After forecasting using RNN/GRU/LSTM models, the forecasted data is inverted to get back to the original condition. Equations 3 and 4 describe the functions used in this study to standardize the dataset.

$$x_t' = (x_t - x_{t-8736}) \qquad\qquad\qquad\qquad\qquad\qquad\qquad (3)$$

$$x_t'' = \frac{x_t' - x_{min}}{x_{max} - x_{min}} \qquad\qquad\qquad\qquad\qquad\qquad\qquad (4)$$

Here, $x_t'$ is the output of the seasonal differencing, $x_t$ is the input time series and $x_{t-8736}$ is the lagged time series by 8736 hours. $x_t''$ is the normalized time series and $x_{min}$ and $x_{max}$ are the minimum and maximum values of the seasonal differenced time series $x_t'$ respectively.

*Baseline Methods*
Before describing the RNN models, the baseline methods need to be discussed. In this study, we have used three baseline models for comparison with RNN based models. The three baseline models are linear regression, ARIMA and HES model. These are popular models for time series forecasting. The linear regression model identifies a straight line to fit the time series based on least square principles. The regression model has been implemented using the "Scikit-learn" package in Python (*23*). ARIMA and HES model is briefly described below.

**ARIMA:** ARIMA stands for Auto-Regressive Integrated Moving Average model. It is a moving average model that also uses seasonal differencing, and forecasts for future time steps using some number of previous time steps in the dataset. This model has three parameters, the lag order, the degree of differencing and the size of the moving average (MA) window. As we are applying seasonal differencing to the time series to make it stationary, we do not require the differencing parameter, so it is set to zero in this study. The lag order and MA window size for each ATR ARIMA model are identified using a grid search method. The values are extracted from the model with the least Akaike Information Criterion (AIC) value (*24*). AIC is used to measure the quality of a statistical model compared to other models. The ARIMA model has been implemented using the "Statsmodels" package in Python (*25*).

**HES:** HES stands for Holt's Exponential Smoothing. It is also known as basic/single exponential smoothing technique or EST model. This smoothing technique can be best explained using equation 5.

$$y_t = \alpha \left( x_t + \sum_{i=1}^{t-1}(1-\alpha)^i x_{t-i} \right) + (1-\alpha)^t x_0 \tag{5}$$

From equation 5, it can be observed that each new prediction ($y_t$) depends on all the previous values ($x_t$) in the time series with continuously increasing powers of coefficients, hence it is called exponential smoothing. Here $\alpha$ is the smoothing factor and it can be anywhere between 0 and 1. A value closer to 1 gives more importance to recent observations in time series, whereas a value closer to 0 gives more importance to smoothing (*26*). The HES model has also been implemented using the "Statsmodels" package in Python (*25*).

*Recurrent Neural Networks*
The predictive model is based on a single RNN layer followed by a dense layer with a single neuron. We have found that this simple architecture performs the best for this dataset. The RNN layer is an individual computation block where the current output is fed back as input for the next step. The structure of a basic RNN model is illustrated using Figure 3. Here, x is the input, y is the output, h is the transfer function and c is the cell state.

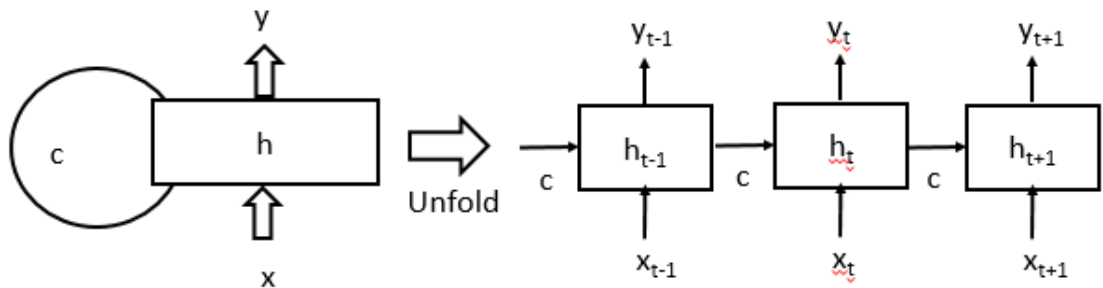

**FIGURE 3 Basic RNN**

Let us assume that $x_t$ is the input, $y_t$ is the output, $h_t$ is the transfer function of the repeater block at time $t$, and $c$ the internal loop in the cell. As a basic RNN cell unfolds, it assumes the form of a chain of cells characterized by a cell-to-cell transfer of vector c. As such, the cell "memorizes" the previous output when computing the next. In this study, we implemented three types of RNN models: Simple RNN, GRU and LSTM. The difference in the RNN models lies in the transfer function (*h*) of the repeater block in Figure 3. The description of each is given below.

**Simple RNN:** In the simple RNN model, the transfer function *h* is merely an activation function. In this study, we have used the "*tanh*" activation function, as shown in Figure 4. The "*tanh*" activation function ensures that the output is normalized between -1 and 1. The simple RNN model operates on two equations, equation 6 and 7. The input to the model is $x_t$ and $c_{t-1}$, and the output of the model is $y_t$ and $c_t$.

$$y_t = tanh(W_y x_t + U_y c_{t-1} + b_y) \tag{6}$$

$$c_t = tanh(W_c x_t + U_c c_{t-1} + b_y) \tag{7}$$

**LSTM:** The Simple RNN problem is characterized by the vanishing gradient problem, preventing the model from determining the long-term internal dependencies. The continuously decreasing gradient of the previous time steps is equivalent to the network forgetting about those time steps. The LSTM model solves the vanishing gradient problem by introducing some additional operations, as shown in Figure 4 (*27*). The top horizontal line, known as the cell state (c), is the memory of the LSTM model, where pointwise additive and multiplicative operations are performed to add or remove information from memory. These operations are the input and forget gates of the LSTM block, which also contains a "*tanh*" activation function for the output. Equations 8-12 describe the mathematical operations inside the LSTM neurons. The input to the model is $x_t$, $c_{t-1}$ and $y_{t-1}$. The output is $y_t$ and $c_t$. Intermediate terms generated inside the neuron are $f_t$, $i_t$ and $o_t$, which correspond to the output of the forget, input and output gates respectively. *W*, *U* and *b* are the weights corresponding to the corresponding gates in the neuron. The dots (.) in the equation indicate elementwise multiplication instead of matrix multiplication. All the terms in the equations can be found in Figure 4.

$$f_t = \sigma(W_f x_t + U_f y_{t-1} + b_f) \tag{8}$$

$$i_t = \sigma(W_i x_t + U_i y_{t-1} + b_i) \tag{9}$$

$$o_t = \sigma(W_o x_t + U_o y_{t-1} + b_o) \tag{10}$$

$$c_t = f_t . c_{t-1} + i_t . tanh(W_c x_t + U_c y_{t-1} + b_c) \tag{11}$$

$$y_t = o_t . tanh(c_t) \tag{12}$$

**GRU:** The GRU is a simplified version of the more complex LSTM unit that combines the input and forgets gates into a single update gate. It then merges both the cell and hidden states for faster operation (*28*). Equations 13-15 describe the mathematical operations inside the GRU neurons.

The intermediate states of the GRU neuron are $z_t$ and $r_t$ respectively. The terms in the equations can be found in Figure 4.

$$z_t = \sigma(W_z x_t + U_z y_{t-1} + b_z) \tag{13}$$

$$r_t = \sigma(W_r x_t + U_r y_{t-1} + b_r) \tag{14}$$

$$y_t = z_t . y_{t-1} + (1 - z_t) . tanh(W_y x_t + U_y(y_{t-1} . r_t) + b_y) \tag{15}$$

The basic diagram of GRU is shown in Figure 4.

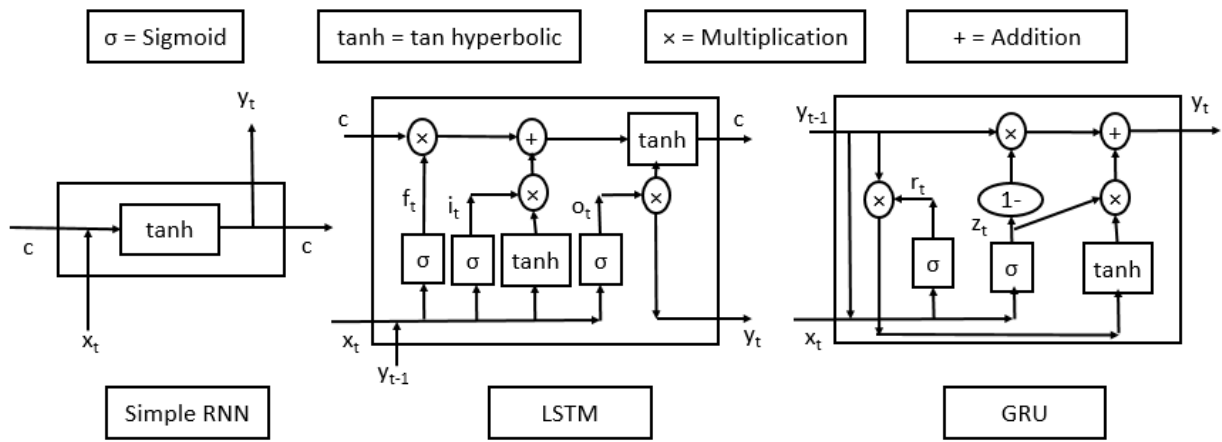

**FIGURE 4 Repeater block transfer function for each model**

A dense layer with one neuron follows up the RNN layer. Here we use a loss function and a gradient descent optimizer to find the optimum weights for this network. All the models are implemented using Keras (*29*) in Python and a Tensorflow (*30*) backend.

**Step 3: Model Evaluation**
After training the model with the training data, we input the testing data. The model predicts the output, and then we compare the two outputs by calculating the errors for hourly traffic volume and AADT prediction. The error block segments the data into blocks of 365 or 366 days based on the prediction year, sums all the hourly volumes and then divides the sum by the total number of days. Two types of error measures are used to compare the models, root mean square error (RMSE) and mean absolute percentage error (MAPE). The performance of different models is detailed in the "Analysis and Results" section below.

**ANALYSIS AND RESULTS**
The analysis and results section contains multiple steps. The steps are described below.
1. At first, we use the data from six sample ATRs of six different functional classes to develop the models. The ATRs selected for analysis are ATR # 1, 6, 15, 21, 39, and 41. These ATRs are selected as they have sufficient data in the time period of 2008-2017. For each ATR, we collect the data from 2008 to 2017 and train the model using the data from 2008 to 2015. The models are then evaluated using their accuracy to predict the AADT of 2016 and 2017. Model accuracy is the opposite measure of the MAPE values. Equation 16 represents the formula of accuracy of the model. For each ATR, we compare 21 different models based on the type of missing data treatment method and the variety of transfer functions used in the repeater block of the RNN model. Based on the accuracy, the best model is selected. The model is checked for overfitting / underfitting issues.

$$Accuracy = (100 - MAPE)\%$$ (16)

2. The best model is compared with the three baseline methods (linear regression, ARIMA, HES) for all ATRs. These four models are used to predict the hourly traffic volumes and AADT of 2016 and 2017 for all ATRs. The mean and variance of RMSE and MAPE values are reported for different functional classes.

3. A statistical significance test is performed for AADT prediction of all ATRs using the four models. The test indicates if there is any significant difference between the actual mean and variance compared to model mean and variance.

4. Next, we analyze the efficacy of the best model in accurately capturing the trends in the time series between 2016 and 2017. A visual representation is shown on how the model captures the long-term dependencies. A sample ATR is chosen for illustration, which in this case is ATR # 39. The visual comparison is accompanied by the RMSE and MAPE values of hourly volume prediction and AADT prediction.

5. Finally, instead of 1 year look-ahead, we perform a multi-year look-ahead prediction (2-year ahead and 3-year ahead) to verify the performance of the best model. In this step, we focus on only AADT prediction, and we conduct the experiment using a sample ATR, which in this case is ATR # 21.

**RNN Model hyper-parameters**
For each RNN / GRU / LSTM model, we vary the hyper-parameters to identify the best parameter for each model. For each model, we vary two hyper-parameters, the number of epochs and batch size. The values of these hyper-parameters are identified from the model with the lowest validation error. Other hyper-parameters of the models are constant for all ATRs in this analysis. The value of the hyper-parameters used in this study are given below.
- Optimizer: Adam (*31*), Learning Rate: 0.001, $\beta_1 = 0.9$, $\beta_2 = 0.999$
- Loss function: Mean Absolute Error
- Number of hidden layers of RNN/GRU/LSTM: 1
- Number of neurons in hidden layer: 1
- Dropout was not required since there was no overfitting / underfitting issues

**Rural Interstate**
At first, we perform a thorough analysis using a sample ATR located in a rural interstate, in this case it is ATR 15. The results for ATR # 15 for the years 2016 and 2017 are shown in Figures 5 and 6, respectively. The LSTM-Median model has the highest accuracy in 2016, and the LSTM-MICE model has the highest accuracy in 2017. The accuracy of the best models is highlighted with a circle.

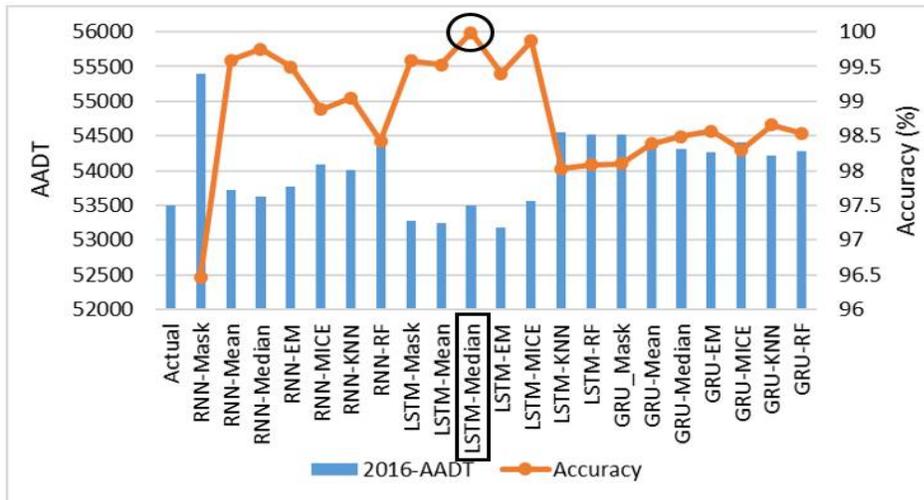

**FIGURE 5 Prediction results for ATR # 15 in 2016**

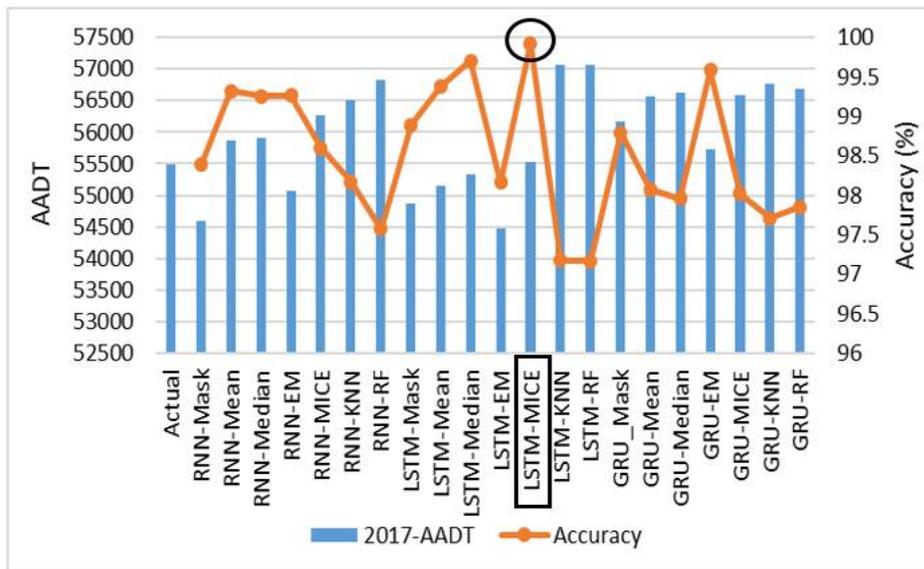

**FIGURE 6 Prediction results for ATR # 15 in 2017**

**Urban Interstate**

For urban interstate, we choose ATR # 21. The AADT prediction of ATR # 21 for the years 2016 and 2017 are shown in Figures 7 and Figure 8, respectively. The LSTM-Median model has the highest accuracy for 2016 and RNN-RF model has the highest accuracy for 2017.

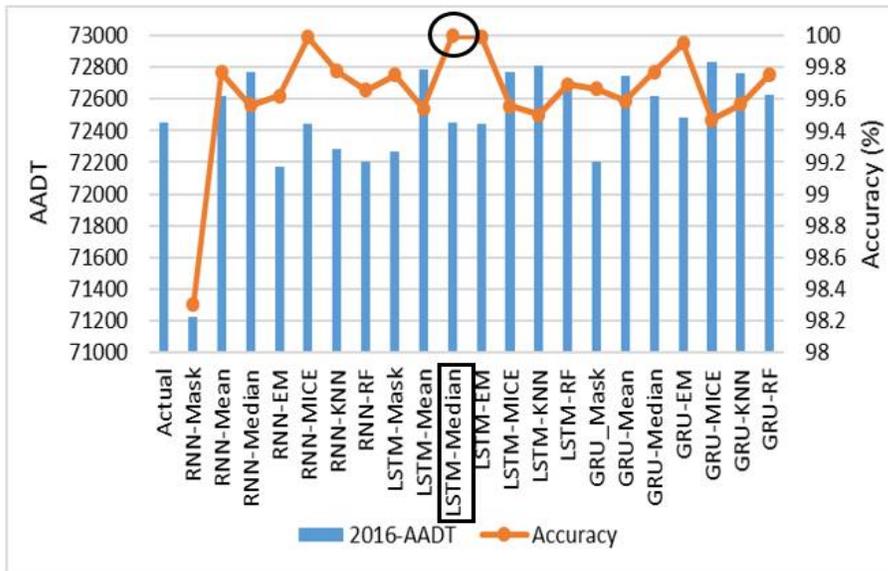

**FIGURE 7 Prediction results for ATR # 21 in 2016**

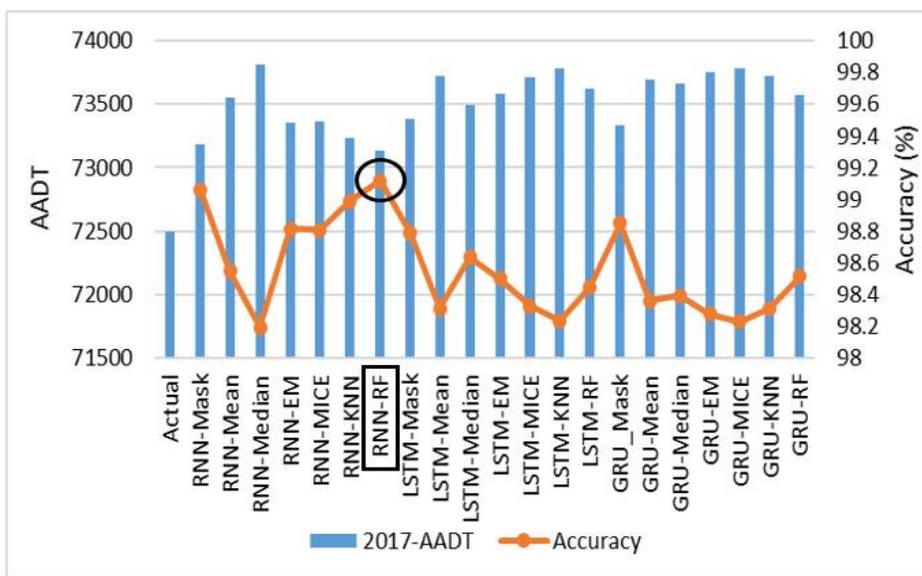

**FIGURE 8 Prediction results for ATR # 21 in 2017**

**Rural Arterial**

For rural arterial, we choose ATR # 1. The AADT prediction of ATR # 1 for the years 2016 and 2017 are shown in Figures 9 and Figure 10, respectively. The LSTM-Median model has the highest accuracy for 2016 and RNN-RF model has the highest accuracy for 2017.

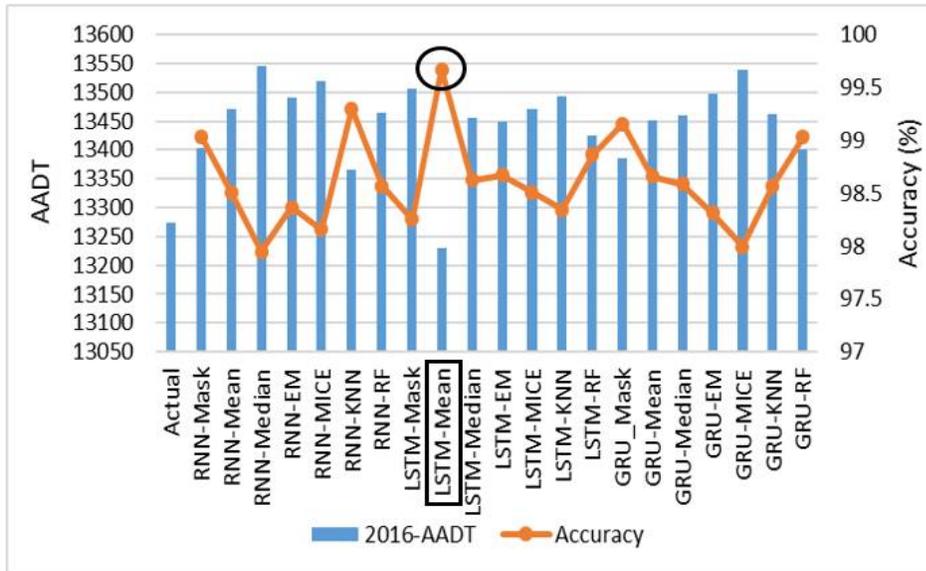

**FIGURE 9 Prediction results for ATR # 1 in 2016**

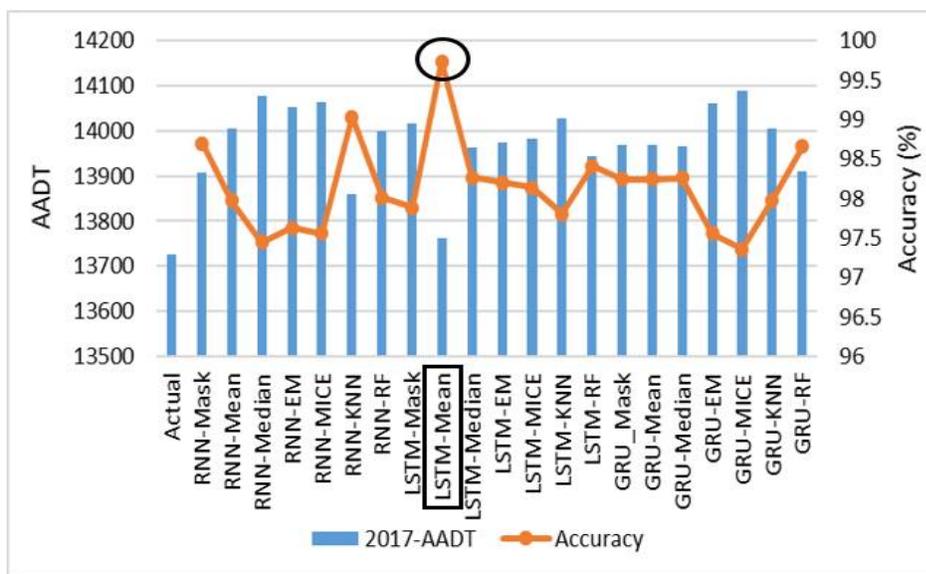

**FIGURE 10 Prediction results for ATR # 1 in 2017**

**Urban Arterial**

For urban arterial, we choose ATR # 6. The AADT prediction of ATR # 6 for the years 2016 and 2017 are shown in Figure 11 and Figure 12, respectively. The GRU-Mean model has the highest accuracy for 2016 and LSTM-KNN model has the highest accuracy for 2017.

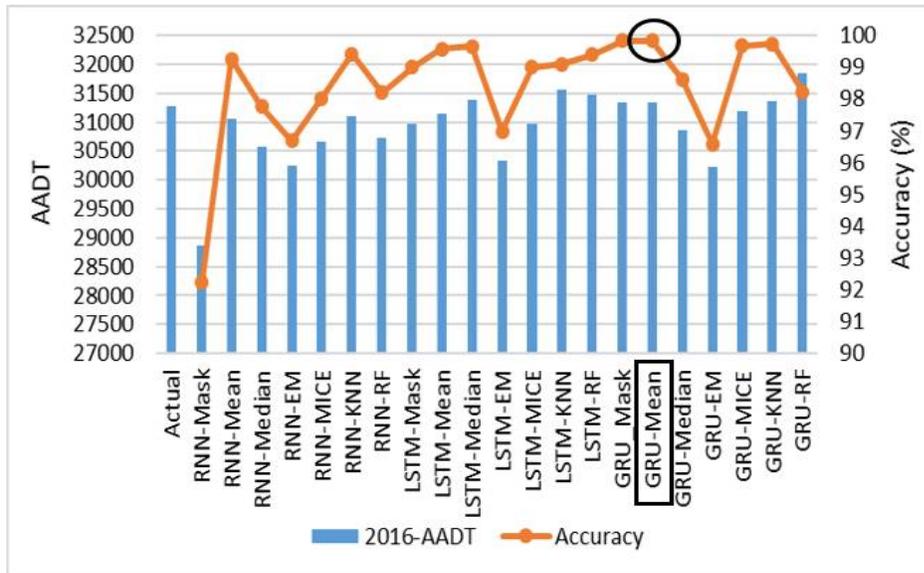

**FIGURE 11 Prediction results for ATR # 6 in 2016**

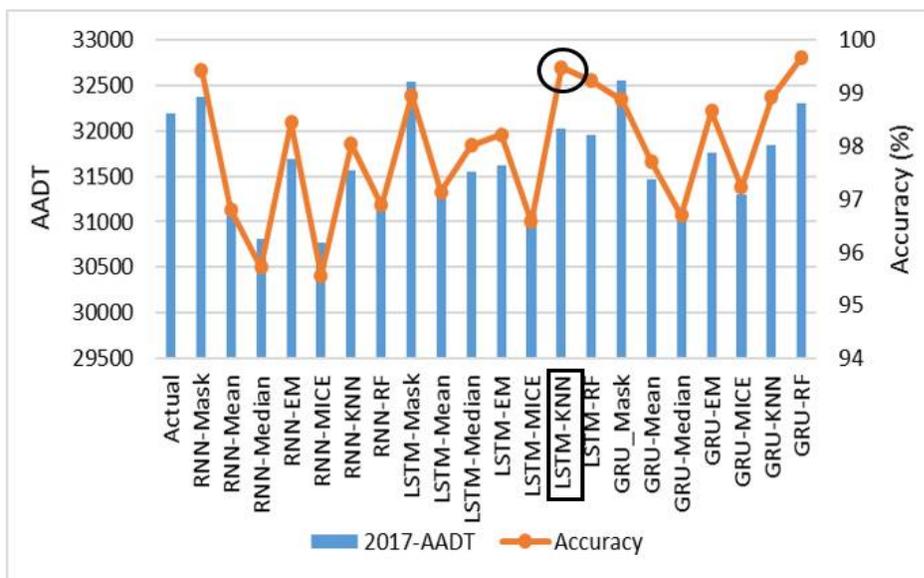

**FIGURE 12 Prediction results for ATR # 6 in 2017**

**Rural Collector**

For rural collector, we choose ATR # 39. The AADT prediction of ATR # 39 for the years 2016 and 2017 are shown in Figures 13 and 14, respectively. The model RNN-KNN has the highest accuracy for 2016 and LSTM-Median model has the highest accuracy for 2017.

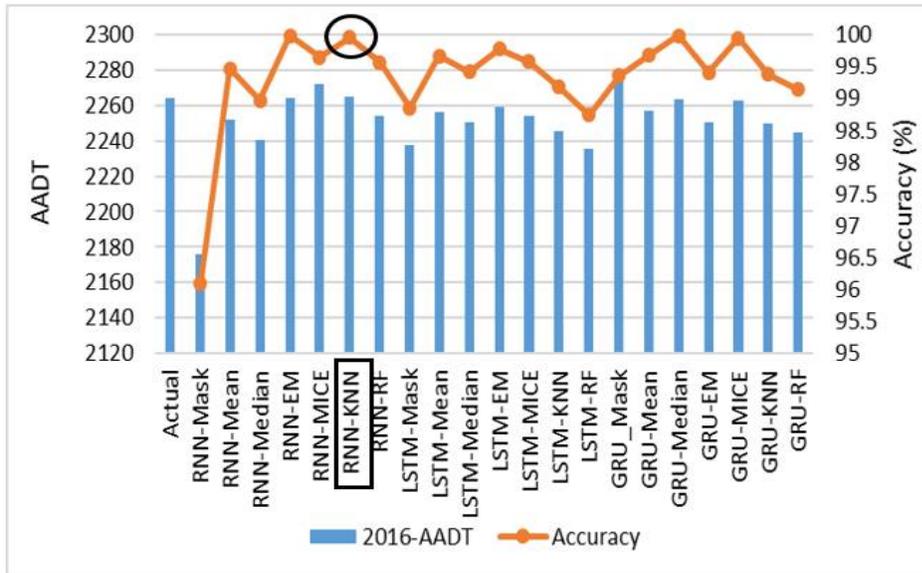

**FIGURE 13 Prediction results for ATR # 39 in 2016**

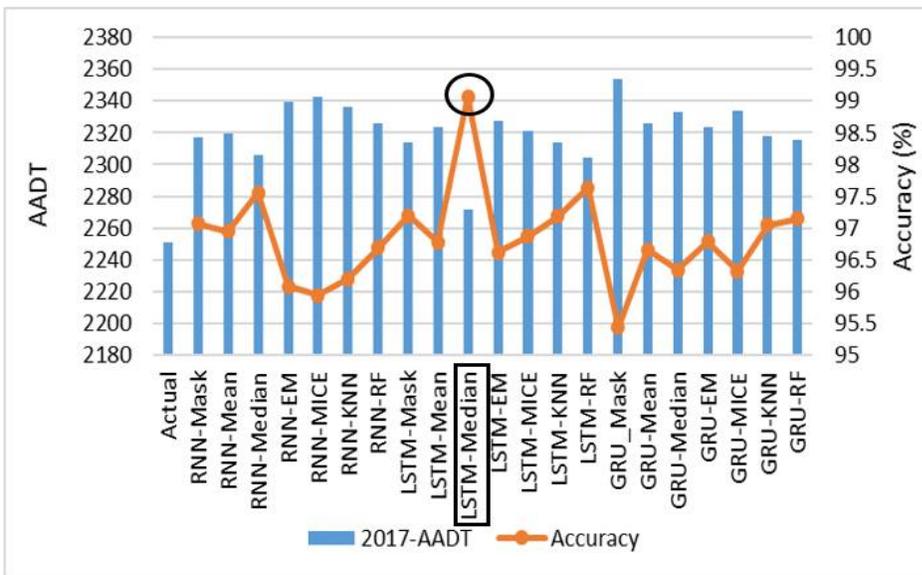

**FIGURE 14 Prediction results for ATR # 39 in 2017**

**Urban Collector**

For urban collector, we choose ATR # 41. The AADT prediction of ATR # 41 for the years 2016 and 2017 are shown in Figures 15 and 16, respectively. The LSTM-KNN model has the highest accuracy for 2016 and LSTM-EM model has the highest accuracy for 2017.

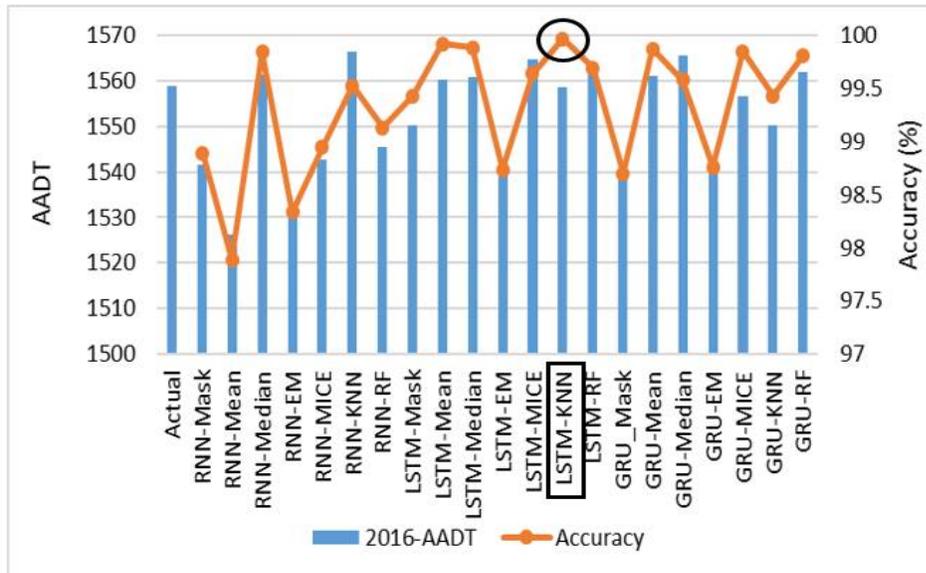

**FIGURE 15 Prediction results for ATR # 41 in 2016**

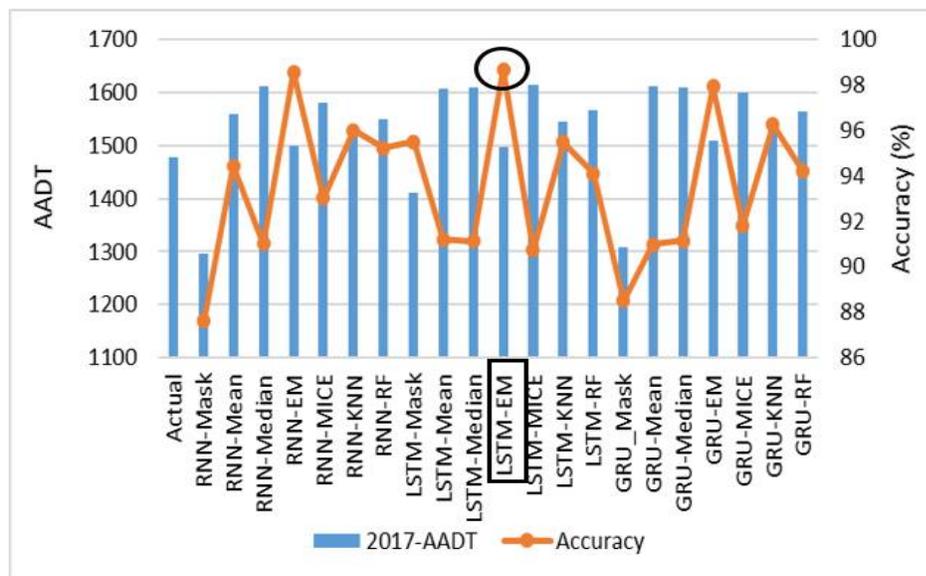

**FIGURE 16 Prediction results for ATR # 41 in 2017**

**Comparison of different RNN models**

In the analysis of the performance of different models, we find that all RNN based models show comparable performance based on average RMSE and MAPE of AADT prediction, LSTM-Median model is chosen for all roadway functional groups. The performance of the LSTM model indicates a significant dependency of the hourly volume on the past data, which the simple RNN model is unable to capture properly. Different imputation methods work well for different functional groups, and imputation performs better than masking in all cases, suggesting that the model struggles to learn from the missing data. Consequently, imputation rather than masking is the preferred method. The performance of the imputation methods depends on the extent of missing data. The percentage of missing values for the ATRs are given in Table 1. Due to the higher percentage of missing values in ATR # 6 and ATR # 41, the more sophisticated KNN method and EM method work better. Concerning the lower percentage of missing values in ATR #1, #15, #21, #39, however, the simple imputation methods of mean and median perform better. This finding suggests that filling the missing values with mean/median is a reasonable method of prediction using the LSTM model, if the percentage of missing values is low. The performance of the LSTM model indicates a significant dependency of the hourly volume on the past data.

**TABLE 1: Missing data (%)**

| ATR | Missing data (%) |
|-----|------------------|
| 1   | 1.3              |
| 6   | 4.9              |
| 15  | 2.2              |
| 21  | 2.9              |
| 39  | 2.1              |
| 41  | 6.7              |

**Overfitting / Underfitting issues**

Based on the analysis in the previous section, the LSTM-Median model is superior with respect to other models. However, we need to verify that there are no overfitting or underfitting issues with the model. Therefore, we check the loss function (MAE) values of the LSTM-Median model for training and validation set for the six ATRs which have been analyzed in this study. The values of the loss MAEs are given in Table 2. As it can be observed, the overfit does not exceed 25% in any case. For ATR # 39, the model actually has a better validation MAE value compared to training MAE, hence the negative sign in the overfit percentage. The maximum overfit is 24% for ATR # 15 and ATR # 41.

**TABLE 2: Training and Validation MAE**

| ATR # | Training MAE | Validation MAE | Overfit (%) |
|-------|--------------|----------------|-------------|
| 1     | 0.020        | 0.023          | 15%         |
| 6     | 0.032        | 0.033          | 3%          |
| 15    | 0.029        | 0.036          | 24%         |
| 21    | 0.021        | 0.025          | 19%         |
| 39    | 0.026        | 0.025          | -4%         |
| 41    | 0.042        | 0.052          | 24%         |

**Comparison with Baseline Models**
As the LSTM-Median model has been selected as the best model, we will perform a comparative study of this model compared to the three baseline models, regression, ARIMA, and HES models. In this study, we predict the hourly traffic volume and AADT using all four models for all ATRs. Then, we calculate the RMSE and MAPE values for all ATRs using all four models. Then, we calculate the mean and standard deviation (SD) of RMSE and MAPE values. As mentioned previously in the method section, ATRs with greater than 20% missing value over the 10 year period of 2008-2017 are not considered in the analysis. Out of 166 ATRs, only 92 ATRs have missing value percentage less than 20%. Table 3 contains all the results for the comparison of LSTM-Median model and the baseline models.

**TABLE 3: RMSE and MAPE values of baseline and LSTM model**

| Model | Statistical Measure | Missing Value (%) | Hourly Volume RMSE | Hourly Volume MAPE (%) | AADT RMSE | AADT MAPE (%) |
|---|---|---|---|---|---|---|
| Regression | Mean | 9.03 | 912 | 157.03 | 1800 | 5.62 |
| | SD | 8.26 | 704 | 82.07 | 1517 | 3.43 |
| ARIMA | Mean | 9.03 | 412 | 25.21 | 1188 | 3.45 |
| | SD | 8.26 | 324 | 10.37 | 1191 | 2.52 |
| HES | Mean | 9.03 | 414 | 29.70 | 1337 | 3.50 |
| | SD | 8.26 | 326 | 13.15 | 1745 | 3.07 |
| LSTM | Mean | 9.03 | 274 | 18.91 | 824 | 2.10 |
| | SD | 8.26 | 216 | 6.55 | 933 | 2.05 |

From Table 3, it can be observed that the LSTM model has the least RMSE and MAPE for both hourly traffic volume and AADT prediction. Linear regression has the highest RMSE and MAPE values, as it is the simplest model. Due to the high mean (9.03%) and standard deviation (8.26%) of missing values in different ATRs, the mean and standard deviation of RMSE and MAPE values of different models have high values for hourly traffic volume prediction. Moreover, the missing values are responsible for high RMSE and MAPE values for hourly traffic volume prediction. The LSTM-Median model has an average accuracy of 81.09% for hourly traffic volume prediction and 97.9% for AADT prediction. The best baseline model is the ARIMA model which has an average accuracy of 74.79% for hourly traffic volume prediction and 96.55% for AADT prediction. The developed model has improved the prediction accuracy from the baseline models.

**Statistical Significance Test of difference**
In the previous subsection, we have shown that based on RMSE and MAPE values, the LSTM-Median model is superior compared to the baseline models. However, in order to prove that there is no significant difference between the model prediction and actual value, we perform a statistical significance test for all four models. We calculate the AADT for 2016 and 2017 for all ATRs using four models. At first, we perform an F-test to identify if the variances are equal or not. Then, we perform a t-test to identify if the sample means are equal. The results of the statistical tests are given in Table 4.

**TABLE 4: Test of significant difference**

|  | F-test | | t-test | |
|---|---|---|---|---|
| Model | p-value | Diff. in Variance | p-value | Diff. in Mean |
| Regression | 0.53 | Insignificant | 0.69 | Insignificant |
| ARIMA | 0.48 | Insignificant | 0.92 | Insignificant |
| HES | 0.53 | Insignificant | 0.88 | Insignificant |
| LSTM | 0.51 | Insignificant | 0.94 | Insignificant |

From Table 4, it can be observed that all four models predict AADT with no significant difference with the actual values. However, the RMSE and MAPE values indicate that the LSTM model is better compared to other models. For example, the regression model may have reasonable accuracy in AADT prediction, but it is not able to capture the high resolution seasonal variations in the traffic volume dataset. Therefore, the LSTM model is always better than regression model.

**Long-Term Dependencies**
Regarding hourly traffic volume prediction, the LSTM-Median model performs better than baseline models with an average MAPE of 18.91%. We next analyze the hourly volume prediction and the model's ability to track the variations in the time series. We compared the actual hourly trend and the predicted hourly trend using the LSTM-Median model for ATR # 39. The comparison is shown in Figure 17.

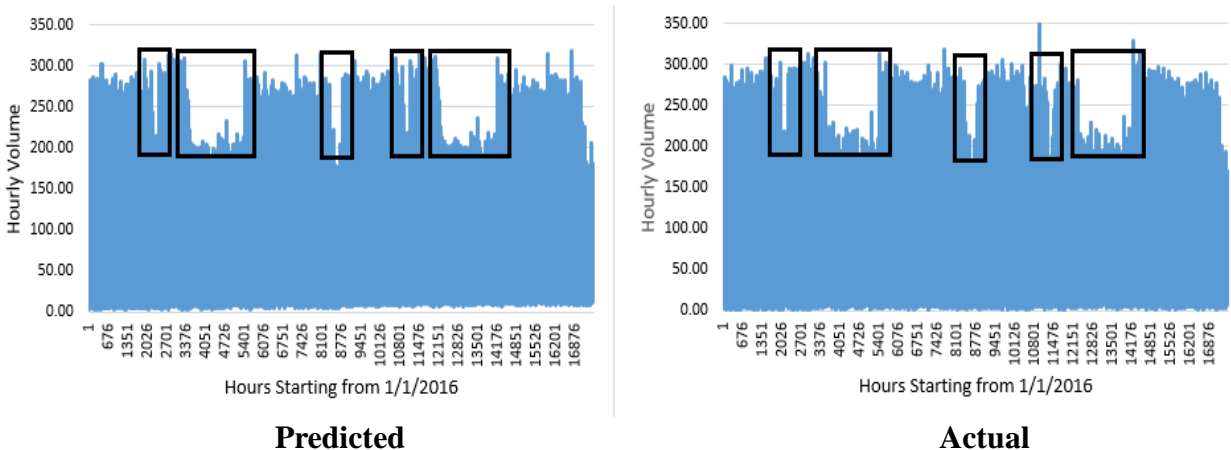

**Predicted**                                     **Actual**

**FIGURE 17 Hourly volume trend (Predicted vs. Actual) for ATR # 39**

The plot on the right side in Figure 17 shows five distinct transitions (highlighted with square boxes) in the actual hourly volume trend. The plot on the left side in Figure 17 shows the effect of the LSTM-Median model in capturing all five transitions (also highlighted with square boxes) accurately. These are seasonal variations that have been observed by the model previously. This indicates its ability to remember the variations using the LSTM network. These are long-term dependencies which are difficult to capture using simple time series prediction models. In terms of hourly traffic volume prediction, the RMSE is 16 and the MAPE is 22.29%. In terms of AADT prediction, the RMSE is 42 and the MAPE is 1.36%. Based on the qualitative and quantitative analysis, it can be concluded that the model can indeed capture the long-term variations in the dataset while maintaining model accuracy.

**Multi-year ahead prediction**
The LSTM-Median model can be used for AADT prediction in the next year. However, the model can be modified to predict further into the future. We conduct a separate experiment to prove the efficacy of the model for 2-year ahead and 3-year ahead AADT forecasts. For this step, we choose the LSTM-median model and the ATR # 21. All the modeling steps remain the same, except the shift operation. Now we shift the data by 730 days for 2-year ahead forecast, and 1095 days for 3-year ahead forecast. Let us first present the case of 2-year ahead forecast. We are predicting the AADT for 2016 and 2017, but we will not be able to use the data from 2015 as input when we are predicting the hourly volumes of 2016. The model loses one year of data due to the look-ahead scenario. Similarly, for the 3-year ahead forecast, the model loses two years of data.

Table 5 shows the impact of increasing the time horizon of the prediction. For 1-year ahead forecast, the LSTM-Median model achieved 99.99% accuracy for 2016 and 98.62% accuracy for 2017. When we perform a 2-year forecast, the accuracy drops to 99.23% and 98.01% respectively. Finally, for 3-year prediction, the accuracy drops further to 98.04% and 97.78% respectively. However, we are still able to achieve accuracies close to 98% despite the lack of data from adjacent years. Therefore, this model is effective in predicting AADT into the future.

**TABLE 5 Multi-Year ahead forecast with LSTM-Median model for ATR 21**

| Model | AADT | Accuracy (%) |
|---|---|---|
| Actual (2016) | 72450 | |
| 1 Year | 72457 | 99.99 |
| 2 Years | 73008 | 99.23 |
| 3 Years | 71028 | 98.04 |
| Actual (2017) | 72500 | |
| 1 Year | 73500 | 98.62 |
| 2 Years | 73892 | 98.01 |
| 3 Years | 70844 | 97.78 |

**CONTRIBUTION OF THIS RESEARCH**
The major contribution of this research is the development of a novel RNN-based predictive model for high accuracy AADT and hourly volume prediction. We have shown that the developed model is capable of capturing the long-term variations in the dataset and in tracking seasonal variations. We also use this model to address the issue of missing data using two approaches. We perform a comparative study and identify LSTM with imputation as the best strategy for high accuracy AADT prediction. To the best of our knowledge, this is the first study involving the development and evaluation of the RNN-based predictive model for future hourly volumes and AADT prediction with missing value treatments. Also, the subsequent investigation of multiple roadway functional groups identifies the best model for each group. Using 92 ATRs, the model is compared with baseline models, achieving the lowest RMSE and MAPE. The hourly traffic volume forecast is accurate and it can capture the long-term variations. Finally, results of the multi-year ahead forecast scenario determine that the model maintains a reasonable accuracy for variable time horizons.

## CONCLUSIONS AND FUTURE STUDY

This study presents the development of the model for AADT prediction that captures the long-term dependencies despite having missing value in the historical dataset. This study also identify the LSTM-Median model as the best model overall for accurately predicting AADT while capturing the long-term seasonal variations in the time series. Overall, the LSTM-Median model is deemed the best model in all scenarios for AADT prediction, with an average RMSE of 274 and MAPE of 18.91% for AADT prediction and average RMSE of 824 and MAPE of 2.10% for hourly traffic volume prediction. The model can capture the long-term variations in the dataset while maintaining high accuracy of AADT and hourly traffic volume prediction. The model is also capable of multi-year ahead forecast with trivial reduction in accuracy.

The input data used here is ATR-specific. In future work, we can investigate more complex model architectures based on RNN units that can predict the hourly volumes and AADT for any functional group. We can train this model on a larger ATR dataset from all ATRs to ensure predictions based on the pattern of the input time series. Finally, we can expand upon the results of this study in AADT prediction modeling to investigate the use of other RNN models (e.g., Elman RNN, Jordan RNN, and Recurrent Multilayer Perceptron networks) and other imputation methods (e.g., cubic spline interpolation, moving average models, and Kalman filters) to develop models with higher accuracy. Moreover, we can investigate novel deep learning architectures other than RNN for hourly volume and AADT prediction. Only ATRs with missing value percentage of less than 20% have been used in this analysis, future models can predict AADT and hourly traffic volume with higher accuracy for all ATRs regardless of the percentage of missing data.

## ACKNOWLEDGMENT

The authors acknowledge the South Carolina Department of Transportation, which provided funding for this research.

## DISCLAIMER

The contents of this report reflect the views of the authors who are responsible for the facts and the accuracy of the presented data. The contents do not reflect the official views of SCDOT or FHWA. This report does not constitute a standard, specification, or regulation.

## AUTHOR CONTRIBUTION STATEMENT

The authors confirm the contribution to the paper as follows:
1. Study conception and design: Zadid Khan, Sakib Mahmud Khan, Mashrur Chowdhury;
2. Data collection: Zadid Khan, Sakib Mahmud Khan;
3. Analysis and interpretation of results: Zadid Khan, Sakib Mahmud Khan, Mashrur Chowdhury;
4. Draft manuscript preparation: Zadid Khan, Sakib Mahmud Khan, Mashrur Chowdhury, Kakan Dey.

All authors reviewed the results and approved the final version of the manuscript.